\documentclass[final,3p,twocolumn]{elsarticle}

\usepackage{graphicx}
\usepackage{float}
\usepackage{amsmath}
\usepackage{amssymb}
\usepackage{amsfonts}
\usepackage[printonlyused]{acronym}
\usepackage[capitalise]{cleveref}
\crefformat{equation}{#2(#1)#3}
\Crefformat{equation}{#2(#1)#3}
\crefname{appendix}{}{}
\usepackage{url}
\usepackage{tikz}
\usetikzlibrary{calc}

\newcommand{\positiontextbox}[4][]{%
  \begin{tikzpicture}[remember picture,overlay]
    \node[inner sep=3pt, fill=yellow,align=left,draw,line width=1pt,#1] at ($(current page.north west) + (#2,-#3)$) {\parbox{.80\paperwidth}{#4}};
  \end{tikzpicture}%
}
\providecommand{\expe}[1]{\ensuremath{\mathrm{e}^{#1}}}
\def\sinc{\mathop{\mathrm{sinc}}\nolimits}

\journal{Signal Processing}

\begin{document}

\begin{frontmatter}
	
	
	
	\title{Time-Modulated Array Beamforming with Periodic Stair-Step Pulses}
	
	\tnotetext[t1]{Declarations of interest: none.}
	
	\tnotetext[t2]{This work has been funded by the Xunta de Galicia (ED431C 2016-045, ED431G/01), the Agencia Estatal de Investigación of Spain (TEC2016-75067-C4-1-R) and ERDF funds of the EU (AEI/FEDER, UE).}
	
	
	\author{Roberto Maneiro-Catoira}
	\ead{roberto.maneiro@udc.es}
	\author{Julio Br\'egains}
	\ead{julio.bregains@udc.es}
	\author{Jos\'e A. Garc\'ia-Naya\corref{cor1}\fnref{fn1}}
	\ead{jagarcia@udc.es}
	\author{Luis Castedo\fnref{fn2}}
	\ead{luis@udc.es}
	\cortext[cor1]{Corresponding author: Jos\'e A. Garc\'ia-Naya, email: jagarcia@udc.es, phone: +34 881016086, postal address: Facultad de Inform\'atica, Campus de Elvi\~na s/n, 15071 A Coru\~na, SPAIN.}
	\address{Universidade da Coru\~na (University of A Coru\~na), CITIC Research Center\\
		Campus de Elvi\~na s/n, 15071 A Coru\~na, Spain}
	\fntext[fn1]{EURASIP member.}

	\begin{abstract}
	Time-modulated arrays (TMAs) are able to improve the side-lobe level of the radiation pattern at the fundamental mode but cannot steer the beam at such a mode towards a given direction. Beam-steering is possible in a TMA, but only at the harmonic patterns and at the expense of a severe TMA efficiency reduction. In this work we propose a TMA approach that simultaneously performs both features over the same beam by using two sets of switches: (1) single-pole four-throw switches to generate periodic stair-step pulses suitable for efficiently synthesizing a uniform steerable beam over the first positive harmonic, and (2) single-pole single-throw switches to reconfigure the side-lobe level of the previous beam. Performance, small size, cost-effectiveness, and performance invariability with the carrier frequency are features that make this TMA approach a competitive solution for analog beamforming. Accordingly, the structure is an attractive proposal for the design of multibeam transceivers.
	\end{abstract}

	\begin{keyword}
	antenna arrays\sep time-modulated arrays\sep antenna efficiency\sep beam steering.
	\end{keyword}

\end{frontmatter}


\positiontextbox{11cm}{28cm}{\footnotesize \textcopyright\ 2020 This manuscript version is made available under the CC-BY-NC-ND 4.0 license \url{https://creativecommons.org/licenses/by-nc-nd/4.0}. This version of the article: Maneiro-Catoira, R., Brégains, J., García-Naya, J. A., \& Castedo, L. (2020). ``Time-modulated array beamforming with periodic stair-step pulses'', has been accepted for publication in Signal Processing, 166(107247). The Version of Record is available online at 
\url{https://doi.org/10.1016/j.sigpro.2019.107247}}

\acrodef{AWGN}[AWGN]{additive white Gaussian noise}
\acrodef{ASK}[ASK]{amplitude-shift keying}
\acrodef{BER}[BER]{bit error ratio}
\acrodef{BFN}[BFN]{beamforming network}
\acrodef{DAC}[DAC]{digital-to-analog converter}
\acrodef{DoA}[DoA]{direction of arrival}
\acrodef{DC}[DC]{direct current}
\acrodef{DSB}[DSB]{double side band}
\acrodef{FSK}[FSK]{frequency-shift keying}
\acrodef{FT}[FT]{Fourier transform}
\acrodef{ISI}[ISI]{inter-symbol interference}
\acrodef{mmWave}[mmWave]{millimeter wave}
\acrodef{MMIC}[MMIC]{monolithic microwave integrated circuit}
\acrodef{NPD}[NPD]{normalized power density}
\acrodef{PSK}[PSK]{phase-shift keying}
\acrodef{QAM}[QAM]{quadrature amplitude modulation}
\acrodef{RF}[RF]{radio frequency}
\acrodef{SA}[SA]{simulated annealing}
\acrodef{SLL}[SLL]{side-lobe level}
\acrodef{SPMT}[SPMT]{single-pole multiple-throw}
\acrodef{SR}[SR]{sideband radiation}
\acrodef{SNR}[SNR]{signal-to-noise ratio}
\acrodef{SPDT}[SPDT]{single-pole double-throw}
\acrodef{SPST}[SPST]{single-pole single-throw}
\acrodef{SP4T}[SP4T]{single-pole four-throw}
\acrodef{SSB}[SSB]{single-sideband}
\acrodef{SWC}[SWC]{sum-of-weighted-cosines}
\acrodef{TM}[TM]{time modulation}
\acrodef{TM-AFN}[TM-AFN]{time-modulation antenna feeding network}
\acrodef{TMA}[TMA]{time-modulated array}
\acrodef{VGA}[VGA]{variable gain amplifier}
\acrodef{VPS}[VPS]{variable phase shifter}
\section{Introduction}\label{sec:introduction}
Hybrid digital-analog beamforming architectures reduce hardware cost and avoid the high energy consumption exhibited by fully digital solutions \cite{Heath2016,Huang2010,MendezRial2016}. Nevertheless, since the analog part in these hybrid solutions is in general based on \acp{VPS}, there is still room for further improvement not only in terms of cost, but also in terms of phase resolution and insertion losses \cite{Qorvo}. Switched \acp{TMA} \cite{ManeiroCatoira2017_Sensors} constitute an interesting alternative to \acs{VPS}-based analog \acp{BFN}.

Although the theoretical basis of the beam-steering capabilities of \acp{TMA} was stated by the pioneering work in \cite{Shanks1961}, it is not until the arrival of systematic optimization algorithms  when \ac{TMA} beam-steering design is addressed with renewed vigor \cite{Li2009PIER,Tong2010}. \ac{TMA} adaptive beamforming was originally studied in \cite{Li2010} considering the radiation patterns at the fundamental mode (static) and at the first harmonics (steerable). Subsequent works address the \ac{TMA} harmonic beamforming by focusing on the mitigation of interfering signals \cite{Poli2011} and on the exploitation of a key differential feature of \acp{TMA}: the exchange of spatial and frequency diversity \cite{Maneiro2017b,Maneiro2017a,Poli2014,Rocca2014,Tong2012}. 
 
A common denominator among all the previous works is that the efficiency of the technique (either beam-steering or beamforming) is compromised by two issues intrinsic to the \ac{TMA} technique \cite{Maneiro2018}:
\begin{enumerate} 
\item The frequency behavior of conventional rectangular pulses, which are not the best ones to efficiently distribute the spectral energy among the multiple harmonic patterns to be exploited \cite{ManeiroCatoira2017_Sensors,Maneiro2017a}.

\item The duplicated-specular radiation diagrams, which are a consequence of the presence of negative harmonics with the same magnitude and opposite phase. Additionally, the scanning inability of the fundamental mode beam and the proportionality between the phases of harmonics with different order apparently jeopardize the versatility of beamforming with \acp{TMA} \cite{Maneiro2018}. A solution to the first issue is proposed in \cite{Bogdan2016} by using \ac{SPDT} switches ---controlled by rectangular sequences with a duty cycle of $50\%$--- and fixed $180^{\circ}$ phase shifters. Such a structure generates only odd harmonics and removes the fundamental pattern, but does not solve the problem of the mirror-frequency diagrams. 

A more efficient beam-steering architecture (with two \ac{SPST} switches per antenna and some additional hardware such as one-bit phase shifters) is proposed in \cite{Amin_Yao2015} to overcome the two aforementioned issues. Such a \ac{SSB} time-modulated phased array architecture synthesizes a steerable uniform beam in the first positive harmonic, and is capable of removing the third-order harmonics, being the highest undesired harmonic that of fifth order. However, the amplitude of such a uniform radiated pattern is still not reconfigurable.
\end{enumerate}
Hence, if we want to reconfigure the amplitude pattern topology (e.g., in terms of \ac{SLL}), while overcoming the aforementioned issues, non-switched  \ac{TMA} beamforming  solutions provide excellent levels of power efficiency, but at the expense of increasing their complexity and, above all, with the handicap of the hardware implementation at high frequencies \cite{Maneiro2018}. 
 
The motivation of this paper is the innovation in the design of switched \acp{TMA} towards the following directions:
\begin{enumerate} 
\item The proposal of an \ac{SSB} switched \ac{TMA} scheme ---based on the novel application of stair-step periodic pulses--- to efficiently perform both beam-steering and beamforming, thus showing the ability of reconfiguring the radiated pattern amplitude.

\item The proposal of a hardware structure for such a switched \ac{TMA} scheme, attractive in terms of cost, size, and complexity.

\item The application of the proposed structure to multibeam transceivers employing a single radio-frequency front-end and showing a performance conditioned by the signal bandwidth rather than the  carrier frequency (as in the case of beamformers based on \acp{VPS} or non-switched \acp{TMA}).
\end{enumerate}

\begin{figure}[!t]
\centering
\includegraphics[width=\columnwidth]{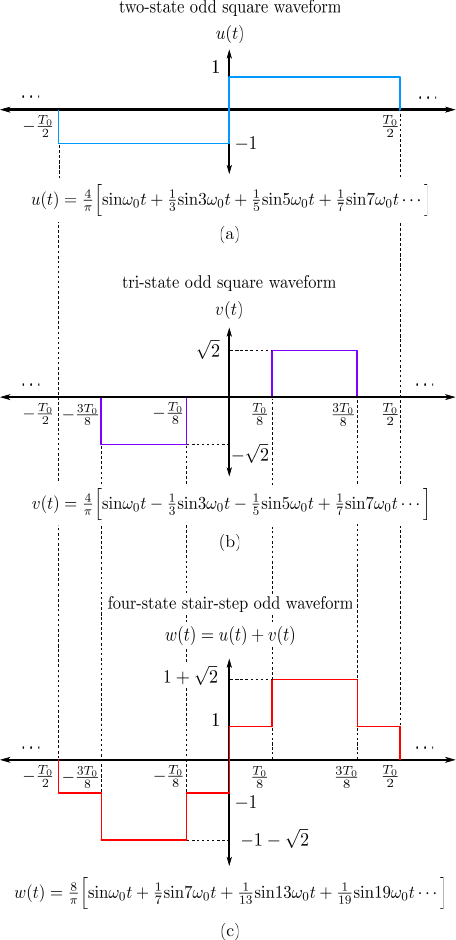}
\caption{Periodic odd waveforms: (a) two-state square waveform with only odd harmonic components, (b) tri-state odd square periodic waveform which inverts the phase of the two harmonics in the middle out of every group of four harmonics, and (c) the sum of the previous waveforms that removes the aforementioned harmonics.}
\label{fig:odd waveforms}
\end{figure}

\section{Characterization of Periodic Stair-Step Pulses}\label{sec:mathematicalAnalysis}
Since our objective is to efficiently exploit the first positive harmonic of a \ac{TMA}, odd periodic rectangular pulses are particularly suitable (apart from its easy and cost-effective implementation) for the following reasons:
\begin{itemize}
	 \item Their \ac{DC} component is zero. Hence, the limitation of having radiation patterns with no steering capability, such as the fundamental one centered at the carrier frequency $\omega_c$, is overcome \cite{Maneiro2017a}.
	 \item Their even harmonic components are also zero, thus providing an advantageous starting point in terms of efficiency with respect to other alternatives.
\end{itemize}
In the following, we look into more detail at these properties. Our point of departure is the simplest odd periodic pulse: the bipolar two-state square waveform $u(t)$ plotted in \cref{fig:odd waveforms}a. The Fourier coefficients of the periodic extension of such a signal, with fundamental period $T_0$, are given by
\begin{align}\label{eq:Fourier coefficients u(t)}
U_{q}=\frac{1}{T_0}\int_{-T_0/2}^{T_0/2}u(t)\expe{-jq\omega_0t}=\begin{cases}
0 & q \hspace{0.2cm}\text{even}\\
-\frac{2j}{\pi q}&q \hspace{0.2cm}\text{odd},
\end{cases}
\end{align}
with $\omega_0=2\pi/T_0$. We observe that the spectral information is entirely contained in the odd harmonics. Notice also that $U_{-q}=U_q^*$, because $u(t)\in \mathbb{R}$. Since the modulus and the argument of $U_q$ are $|U_q|=2/(\pi q)$ and $\sphericalangle U_q=-\pi/2$, respectively, we can easily express the Fourier series expansion of $u(t)$ as (see \cref{fig:odd waveforms}a)
\begin{equation}\label{eq:u(t) sin}
u(t)=\frac{4}{\pi}\sum_{q = 1,3,5, \dots}^\infty\frac{1}{q}\sin(q\omega_0t).
\end{equation}	
Notice that when we apply this kind of pulses to a uniformly excited TMA \cite{Bogdan2016,Bogdan2016a}, we find that the peak level of
the first undesired harmonic (the third one) is too high ($-9.54$\,dB) with respect to the first-order harmonic peak, making these pulses very inefficient for our purposes. Hence, it would be desirable for our target pulses to eliminate the third and even the fifth order harmonics. To construct pulses with such features, we turned our attention to the tri-state odd square waveform $v(t)$ in \cref{fig:odd waveforms}b, whose Fourier series coefficients are 
\begin{align}\label{eq:Fourier coefficients v(t)}
V_{q}&=\frac{j\sqrt2}{q\pi}\left[\cos\left(\frac{3\pi q}{4}\right)-\cos\left(\frac{\pi q}{4}\right)\right]\notag\\
&=\begin{cases}
0 & q \hspace{0.2cm}\text{even}\\
(-1)^{\frac{(q+1)(q-1)}{8}}\left(-\frac{2j}{\pi q}\right) & q \hspace{0.2cm}\text{odd},
\end{cases}
\end{align}
in which the sign of $V_q$ is derived in \cref{Appendix:derivation of sign Vq}.
Analogously to the pulse $u(t)$, the Fourier series expansion is easily derived from \cref{eq:Fourier coefficients v(t)}, arriving at (see also \cref{fig:odd waveforms}b)
\begin{equation}\label{eq:v(t) sin}
v(t)=\frac{4}{\pi}\sum_{q = 1,3,5, \dots}^\infty(-1)^{\frac{(q+1)(q-1)}{8}}\frac{1}{q}\sin(q\omega_0t).
\end{equation}
As a result, the pulse $w(t)=u(t)+v(t)$ (see \cref{fig:odd waveforms}c) is a four-state stair-step odd waveform with Fourier series coefficients
\begin{align}\label{eq:Fourier coefficients w(t)}
W_{q}=U_q+V_q=\begin{cases}
-\frac{4j}{\pi q}&\text{$|q|\in\Upsilon$}\\
0 & \text{otherwise},
\end{cases}
\end{align}
with $\Upsilon=\{4\alpha+(-1)^\alpha-2; \alpha\in \mathbb{N^*}\}=\{1, 7, 9, 15, 17, 23,25, 31, \dots\}$, and therefore
\begin{equation}\label{eq:w(t) sin}
w(t)=\frac{8}{\pi}\sum_{q \in \Upsilon}\frac{1}{q}\sin(q\omega_0t).
\end{equation}
Additionally, and in order to steer the exploited \ac{TMA} harmonic pattern, we must consider a time-shifted version of $w(t)$, $w_n(t)= w(t-D_n)$, with $D_n$ being the corresponding time-delay variable, leading to
\begin{equation}\label{eq:w_n(t) sin}
w_n(t)=\frac{8}{\pi}\sum_{q \in \Upsilon}\frac{1}{q}\sin(q\omega_0(t-D_n)).
\end{equation}
In the following sections, we consider the pulse $w_n(t)$ to efficiently design a flexible \ac{SSB} \ac{TMA} beamformer.

\begin{figure*}[t]
	\centering
	\includegraphics[width=0.6\linewidth]{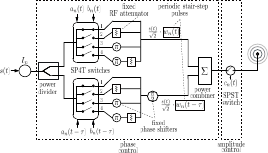}
	\caption{Feeding scheme for the $n$-th antenna element of the proposed
		\ac{SSB} \ac{TMA} beamformer, being $w_n(t)$ a periodic stair-step pulse, $c_n(t)$ a periodic rectangular pulse, and $\tau$ a time-delay. The phase and the amplitude of the corresponding time-modulated excitation are controlled separately.}
	\label{fig:block diagram n-th element}
\end{figure*}

\section{SSB TMA Beamformer Using Periodic Stair-Step Pulses}\label{sec:SSB TMA with stair-step pulses model}

Let us consider a linear array with $N$ isotropic elements with unitary static excitations $I_n = 1$, $n \in \{0, 1,\dots, N-1\}$, whose $n$-th element feeding scheme is illustrated in \cref{fig:block diagram n-th element}. The phase and the amplitude of the $n$-th dynamic excitation of the array are controlled separately. More specifically,  the phase is controlled with a two-branch structure: one branch is time-modulated by a stair-step periodic ($T_0$) waveform $w_n(t)$, whereas the other one is time-modulated by $w_n(t-\tau)$, with $\tau$ being a previously fixed time delay, followed by a $\pi/2$ fixed phase shifting. On the other hand, the amplitude is controlled by a periodic ($T_0$) rectangular pulse $c_n(t)$ generated by means of an \ac{SPST} switch.

Let us see in detail how $w_n(t)$ and $w_n(t-\tau)$  are applied to the $n$-th array excitation (\cref{fig:block diagram n-th element}).  
The four  levels of the stair-step pulse are generated by means of a \ac{SP4T} switch controlled by the periodic ($T_0$) binary signals $a_n(t)$ and $b_n(t)$ whose states are specified in \cref{tab: control SP4T}. 

\begin{table}[b]
	\caption{Control of an SP4T switch (during one period $T_0$) to time-modulate the array excitations with the stair-step periodic signal $w_n(t)$ in \cref{fig:odd waveforms}c. The position of the \ac{SP4T} switch is determined by the corresponding levels of the switch control periodic ($T_0$) signals $a_n(t)$ and $b_n(t)$.}
	\label{tab: control SP4T}
	\begin{center}
		\setlength\tabcolsep{0.2em}
		\def\arraystretch{1.1}
		\begin{tabular}{|c|c||c|c|}
			\hline
			{$a_n(t)$}&{$b_n(t)$}&{\text{SP4T output}} & {$w_n(t)$}\\
			\hline \hline
			$0$ & $0$ & $1$ & $1+\sqrt2$\\\hline
			$0$ & $1$ & $2$ & $1$\\\hline
			$1$ & $0$ & $3$ & $-1-\sqrt2$\\\hline
			$1$ & $1$ & $4$ & $-1$\\\hline
		\end{tabular}
	\end{center}
\end{table}

More specifically, the switch terminals are connected physically to the antenna according to:
\begin{itemize}
	\item Terminal 1: direct connection. 
	\item Terminal 2: connection through a fixed \ac{RF} attenuator of $-20\text{log}(1+\sqrt2)$\,dB.
	\item Terminal 3: connection through a fixed 180$^\circ$ phase shifter.
	\item Terminal 4: connection through a fixed phase shifter followed by a fixed \ac{RF} attenuator, both with the same characteristics as those employed in Terminal 3 and Terminal 2 branches, respectively.
\end{itemize}
The second \ac{SP4T} switch is used to apply the delayed periodic pulse $w_n(t-\tau)$ to the $n$-th array excitation. Such a switch is controlled by the corresponding delayed versions of $a_n(t)$ and $b_n(t)$.

According to \cref{fig:block diagram n-th element}, the time-varying array factor will be given by
\begin{equation}\label{eq:array factor time-domain}
F(\theta,t)=\sum_{n=0}^{N-1}c_n(t)\left[\frac{w_n(t)}{\sqrt{2}}+j\frac{w_n(t-\tau)}{\sqrt{2}}\right]\expe{j\beta z_n\cos\theta},
\end{equation} 
where $z_n$ represents the $n$-th array element position on the $z$ axis, $\theta$ is the angle with respect to such a main axis, $\beta=2\pi/\lambda$ represents the wavenumber for a carrier wavelength $\lambda = 2\pi \mathrm{c} /\omega_c$, where $\mathrm{c}$ is the speed of light and $\omega_c$ is the carrier frequency of the communication signal $s(t)$ shown in \cref{fig:block diagram n-th element}. Notice that $\omega_c$ is not explicitly included in the array factor in \cref{eq:array factor time-domain} as in previous works \cite{ManeiroCatoira2018_CommLetters,ManeiroCatoira2018b_AWPL}. We begin the analysis of \cref{eq:array factor time-domain} by evaluating the term $w_n(t)+jw_n(t-\tau)$. For the sake of simplicity, we will analyze the \ac{FT} of such a term, i.e., $\mathrm{FT}[w_n(t)]+j\mathrm{FT}[w_n(t-\tau)]$ where, by virtue of \cref{eq:w_n(t) sin}, 
\begin{align}\label{eq:pulses_frequency_domain}
&\mathrm{FT}[w_n(t)]=\notag\\ &=\frac{8}{j}\sum_{q\in\Upsilon}\frac{1}{q}[\expe{-jq\omega_0D_n}\delta(\omega-q\omega_0)-\expe{jq\omega_0D_n}\delta(\omega+q\omega_0)],\notag\\
&\text{and}\notag\\
&\mathrm{FT}[w_n(t-\tau)]=\expe{-j\omega \tau}\mathrm{FT}[w_n(t)]=\notag\\ &=\frac{8}{j}\sum_{q\in\Upsilon}\frac{1}{q}[\expe{-jq\omega_0 \tau}\expe{-jq\omega_0D_n}\delta(\omega-q\omega_0)-\notag\\&-\expe{jq\omega_0 \tau}\expe{jq\omega_0D_n}\delta(\omega+q\omega_0)],
\end{align} 
where $\delta(\omega)$ is the unit impulse in the frequency domain. If we select a delay $\tau$ verifying that $\omega_0\tau=\pi/2$, then $\expe{-jq\omega_0\tau}=(-j)^q$ and $\expe{jq\omega_0\tau}=j^q$, and hence we have that
\begin{align}\label{eq:suma de wn y delayed frequency_domain}
&\mathrm{FT}[w_n(t)]+j\mathrm{FT}[w_n(t-\tau)]=\notag\\
&=\frac{8}{j}\sum_{q\in\Upsilon}\frac{1}{q}\Bigg[(1-(-j)^{q+1})\expe{-jq\omega_0D_n}\delta(\omega-q\omega_0)+\notag\\
&+(-1-j^{q+1})\expe{jq\omega_0D_n}\delta(\omega+q\omega_0)\Bigg].
\end{align}
Considering the sets of indexes $\Upsilon_1=\{8\alpha-7; \alpha\in \mathbb{N^*}\}=\{1,  9, 17, \dots\}$ and $\Upsilon_2=\{8\alpha-1; \alpha\in \mathbb{N^*}\}=\{7, 15, 23, \dots\}$, verifying $\Upsilon=\Upsilon_1\cup\Upsilon_2$, we realize that
\begin{align}
1-(-j)^{q+1}&=\begin{cases}
2 & q\in \Upsilon_1\\
0&q\in \Upsilon_2,
\end{cases};\  \notag\\
-1-j^{q+1}&=\begin{cases}
-2 & q\in \Upsilon_2\\
0 & q\in \Upsilon_1.
\end{cases}
\end{align}
Hence, we rewrite \cref{eq:suma de wn y delayed frequency_domain} as
\begin{align}\label{eq:suma de wn y delayed frequency_domain 2}
&\mathrm{FT}[w_n(t)]+j\mathrm{FT}[w_n(t-\tau)]=\notag\\
&=\frac{16}{j}\sum_{q\in\Upsilon_1}\frac{1}{q}\expe{-jq\omega_0D_n}\delta(\omega-q\omega_0)+\notag\\
&+\frac{(-16)}{j}\sum_{q\in\Upsilon_2}\frac{1}{q}\expe{jq\omega_0D_n}\delta(\omega+q\omega_0),
\end{align}
and we realize that the harmonics with order $-1,7,-13, 19, \dots$ are removed. By applying the inverse \ac{FT} to \cref{eq:suma de wn y delayed frequency_domain 2}, we have
\begin{align}\label{eq:suma de wn y delayed time domain plus}
&w_n(t)+jw_n(t-\tau)=\notag\\
&=\sum_{q\in\Upsilon_1}\frac{8}{j\pi q}\expe{-jq\omega_0D_n}\expe{jq\omega_0t}\notag\\
&+\sum_{q\in\Upsilon_2}\frac{(-8)}{j\pi q}\expe{jq\omega_0D_n}\expe{-q\omega_0t}.
\end{align}
On the other hand, the Fourier series expansion of $c_n(t)$ is given by
\begin{equation}\label{eq:c_n(t)}
c_n(t)=\sum_{k=-\infty}^{\infty}C_{nk}\expe{jk\omega_0t},
\end{equation}
with $C_{nk}=\xi_n\sinc(k\pi\xi_n)\expe{-jk\pi\xi_n}$ \cite{Fondevila2006}. In this expression, $\sinc(x) = \sin(x)/x$, and $\xi_n \in (0,1] \subset \mathbb{R}$ are the normalized pulse time durations. 
By considering \cref{eq:suma de wn y delayed time domain plus} and \cref{eq:c_n(t)}, we can rewrite \cref{eq:array factor time-domain} as
\begin{small}
\begin{align}\label{eq:F time plus}
&F(\theta,t)=\notag\\
&=\frac{1}{\sqrt{2}}\sum_{n=0}^{N-1}\Bigg[\sum_{k=-\infty}^{\infty}C_{nk}\expe{jk\omega_0t}\Bigg(\sum_{q\in\Upsilon_1}\frac{8}{j\pi q}\expe{-jq\omega_0D_n}\expe{jq\omega_0t}\notag\\
&+\sum_{q\in\Upsilon_2}\frac{(-8)}{j\pi q}\expe{jq\omega_0D_n}\expe{-jq\omega_0t}\Bigg)\Bigg]\notag\\
&=\frac{1}{\sqrt{2}}\sum_{k=-\infty}^{\infty}\sum_{q\in\Upsilon_1}\expe{j(k+q)\omega_0t}\notag\\
&\cdot\sum_{n=0}^{N-1}\underbrace{C_{nk}\tfrac{8}{j\pi q}\expe{-jq\omega_0D_n}}_{=(I_n)^k_q}\expe{jkz_n\cos\theta}\notag\\
&+\frac{1}{\sqrt{2}}\sum_{k=-\infty}^{\infty}\sum_{q\in\Upsilon_2}\expe{j(k-q)\omega_0t}\notag\\
&\cdot\sum_{n=0}^{N-1}\underbrace{C_{nk}\tfrac{(-8)}{j\pi q}\expe{jq\omega_0D_n}}_{=(I_n^{'})^k_q}\expe{jkz_n\cos\theta}.
\end{align}  
\end{small}
We now define
\begin{align}\label{eq:definition of F1 y F2}
F_1(\theta,t)^k_q &= \expe{j(k+q)\omega_0t}\sum_{n=0}^{N-1} (I_n)^k_q\cdotp\expe{jkz_n\cos\theta},\notag\\
F_2(\theta,t)^k_q &= \expe{j(k-q)\omega_0t}\sum_{n=0}^{N-1} (I_n^{'})^k_q\cdotp\expe{jkz_n\cos\theta},
\end{align} 
where the dynamic excitations  $(I_n)^k_q$ and $(I_n^{'})^k_q$ are given by
\begin{align}\label{eq:dynamic excitations}
&(I_n)^k_q=\frac{8 C_{nk}}{j\pi \sqrt{2}q}\expe{-jq\omega_0D_n}, \hspace{0.2cm}q\in \Upsilon_1,\hspace{0.2cm}k\in \mathbb{Z},\notag\\
&(I_n^{'})^k_q=\frac{-8 C_{nk}}{j\pi \sqrt{2}q}\expe{jq\omega_0D_n}, \hspace{0.2cm}q\in \Upsilon_2,\hspace{0.2cm}k\in \mathbb{Z},
\end{align}
to finally obtain
\begin{equation}\label{eq:array factor3}
F(\theta,t)=\sum_{k=-\infty}^{\infty}\left[\sum_{q\in\Upsilon_1}F_1(\theta,t)^k_q+\sum_{q\in\Upsilon_2}F_2(\theta,t)^k_q \right].
\end{equation}	
Notice that we must consider the normalized version of the pulses in \cref{fig:odd waveforms}c since we are not considering a voltage gain of $1+\sqrt{2}$ at the input of the \acp{SP4T}. Consequently, the normalized dynamic excitations are
\begin{align}\label{eq:normalized dynamic excitations}
&(\overline{I_n})^k_q=\frac{8C_{nk}}{j\sqrt{2}(1+\sqrt{2})\pi q}\expe{-jq\omega_0D_n}, \hspace{0.2cm}q\in \Upsilon_1,\hspace{0.2cm}k\in \mathbb{Z},\notag\\
&(\overline{I_n^{'}})^k_q=\frac{-8C_{nk}}{j\sqrt{2}(1+\sqrt{2})\pi q}\expe{jq\omega_0D_n}, \hspace{0.2cm}q\in \Upsilon_2,\hspace{0.2cm}k\in \mathbb{Z},
\end{align}
and therefore, $|(\overline{I_n})^k_q|^2=|(\overline{I_n^{'}})^k_q|^2=32|C_{nk}|^2/((1+\sqrt{2})\pi q)^2$. Notice that $C_{nk}$ will be selected in the conventional way of \ac{TMA} design, i.e., using systematic optimization algorithms to maintain the side-lobe zone of the radiated power of the fundamental mode $k=0$ under a certain and previously stipulated level, whereas the radiated power over the undesired harmonics is minimized. Hence, only $C_{n0}=\xi_n$ will be meaningful in the \ac{SSB} \ac{TMA} pattern design. Once $k=0$ is considered, the two most significant $|(\overline{I_n})^k_q|^2$ are those  for $q=1\in \Upsilon_1$ and for $q=7\in \Upsilon_2$, hence satisfying that $20\log|(\overline{I_n})^0_7/(\overline{I_n})^0_1|=-16.9$ dB. Therefore, the aim of the technique is to guarantee a single useful harmonic beam pattern: the one given by   $|F_1(\theta,t)^0_1|^2$.

\section{Efficiency of the Time Modulation}\label{sec:efficiency}
In this section we determine the efficiency of the time modulation operation in the proposed \ac{SSB} \ac{TMA} beamformer. For the sake of simplicity, but without any relevant loss of generality, we will consider a uniform linear array with $\lambda/2$ of inter-element distance transmitting a single carrier with normalized power. Such an efficiency can be split into two
separate efficiencies
\begin{equation}\label{eq:eficiencias}
\eta =\eta_\text{TMA}\cdot \eta_\text{BFN},
\end{equation}
whose interpretations are described below.
\begin{figure*}[h]
	\centering
	\includegraphics[width=13cm]{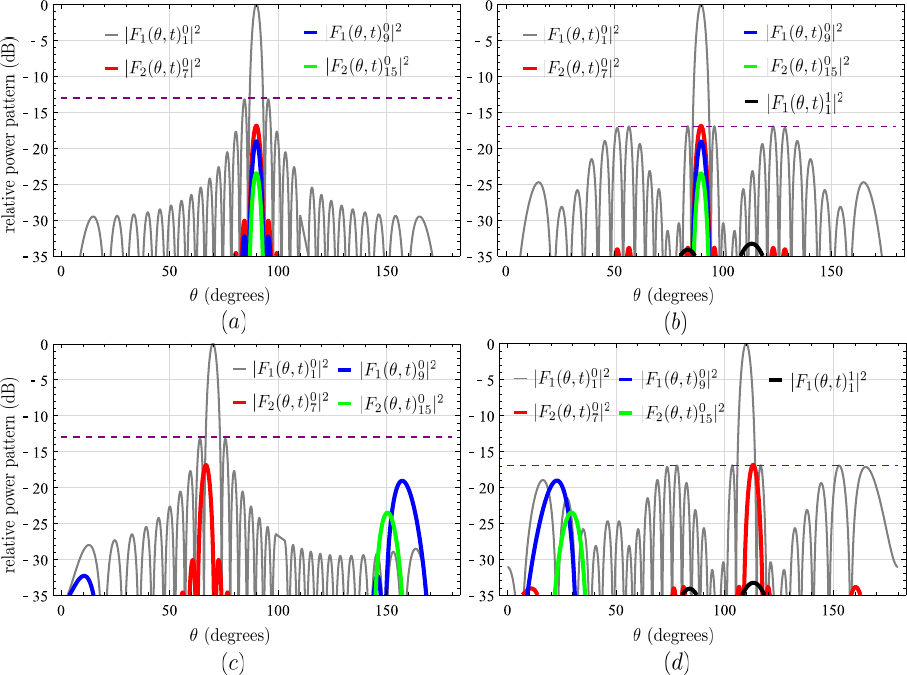}
	\caption{By considering an array of $N=30$ elements treated as the one illustrated in \cref{fig:block diagram n-th element}, we show the normalized power radiated patterns for the following \ac{TMA} configurations: (a) Phased array mode (\ac{SPST} switches closed) with $D_n=0$ ($\theta_{\text{scan}}=90^\circ$). The first undesired harmonic, at $\omega_c-7\omega_0$, has a $-16.90$\,dB peak level. The subsequent harmonics, at $\omega_c+9\omega_0$ and $\omega_c-15\omega_0$, have peak levels of $-19.08$\,dB and $-23.52$\,dB, respectively. The efficiency of the time modulation is $\eta=0.56$ ($-2.55$\,dB). (b) Beamformer mode with $D_n=0$ ($\theta_{\text{scan}}=90^\circ$). The \ac{SPST} switches are governed by periodic sequences with the $\xi_n$ values specified in \cref{tab: xi_n}. The \ac{SLL} of the desired pattern is  now set at $-17$\,dB. Notice that, apart from the undesired harmonics of the phased array mode, the most significant harmonic due to the amplitude time modulation is the one corresponding to $\omega_c+2\omega_0$, which is below $-30$\,dB. The price to be paid for the \ac{SLL} improvement is a certain worsening of the efficiency: $\eta=0.46$ ($-3.37$\,dB). (c) Phased array mode with $D_n$ selected to accomplish a $\theta_{\mathrm{scan}}=70^\circ$. (d) Beamformer mode with $D_n$ selected to accomplish a $\theta_{\mathrm{scan}}=110^\circ$.}
	\label{fig:radiated pattern phased array1}
\end{figure*}
The term $\eta_{\text{TMA}}$ accounts for the ability of the \ac{TMA} technique to radiate only over the useful harmonics. It is determined by
\begin{equation}\label{eq:eta_TMA}
\eta_{\text{TMA}}=\frac{P_U^{\text{TM}}}{P_R^{\text{TM}}},
\end{equation}  where $P_U^{\text{TM}}$ and $P_R^{\text{TM}}$ are the useful and the total mean powers, respectively, radiated by the \ac{SSB} \ac{TMA} beamformer. It is remarkable that the \ac{SSB} operation at least doubles the value of this efficiency with respect to that of a conventional \ac{TMA}.

Most of the works available in the literature analyzing the \ac{TMA} efficiency limit themselves to the study of $\eta_{\text{TMA}}$. The second component of the efficiency, $\eta_{\text{BFN}}$, accounts for the reduction of the total mean power radiated  by a uniform static array caused by the insertion of the \ac{TMA} \ac{BFN}. $\eta_{\text{BFN}}$ is of critical importance due to its high impact on the antenna gain. This efficiency is evaluated by means of the quotient 
\begin{equation}\label{eq:eta_BFN}
\eta_{\text{BFN}}=\frac{P_R^{\text{TM}}}{P_R^{\text{ST}}},
\end{equation}
where $P_R^{\text{ST}}$ is the total mean power radiated by a uniform static array with $N$ elements.

Let us now analyze in detail both efficiencies. We will start by deriving the expression of $\eta_{\text{TMA}}$. In the proposed design, $P_R^{\text{TM}}$ is given by \cite{Maneiro2017a}
\begin{equation}\label{eq:PR TM}
P_R^{\text{TM}}=\sum_{k=-\infty}^{\infty}\left[\sum_{q\in\Upsilon_1}(p_1)^k_q+\sum_{q\in\Upsilon_2}(p_2)^k_q\right],
\end{equation}
being $(p_1)^k_q$ and $(p_2)^k_q$ the mean transmit power values at the harmonics $\omega_c+(k+q)\omega_0$ and $\omega_c+(k-q)\omega_0$, respectively.
Since $(p_1)^k_q=(p_2)^k_q=4\pi \sum_{n=0}^{N-1}|(\overline{I_n})^k_q|^2 $ \cite{Maneiro2017a}, by considering \cref{eq:normalized dynamic excitations}, we have
 \begin{equation}\label{eq:p_1 k}
 (p_1)^k_q=(p_2)^k_q=\frac{128}{\pi(1+\sqrt{2})^2} \sum_{n=0}^{N-1}\xi_n^2\sinc^2(k\pi\xi_n)/q^2.
\end{equation} 
Thus, we can rewrite \cref{eq:PR TM} as
\begin{equation}\label{eq:PR TM 1}
P_R^{\text{TM}}=\frac{128}{\pi(1+\sqrt{2})^2}\sum_{n=0}^{N-1}\sum_{q\in\Upsilon}\sum_{k=-\infty}^{\infty}\frac{\xi_n^2\sinc^2(k\pi\xi_n)}{q^2}.
\end{equation}
Having now in mind that for all $\xi_n\in (0,1]$ the sinc-square infinite series converges to
$\sum_{k=-\infty}^{\infty}\sinc^2(k\pi\xi_n)=1/\xi_n$,
we can express the total mean power as
\begin{equation}\label{eq:PR TM 2}
P_R^{\text{TM}}=\frac{128}{\pi(1+\sqrt{2})^2}\sum_{q\in\Upsilon}\frac{1}{q^2}\sum_{n=0}^{N-1}\xi_n,
\end{equation}
and since the infinite series $\sum_{q\in\Upsilon}1/q^2=1/64\left(\Psi_1(1/8)+\Psi_1(7/8)\right)$, with $\Psi_1$ the polygamma function of order 1 (see \cref{subsec:derivation of the sum in eta_P}), by denoting $A_0=\sum_{q\in\Upsilon}1/q^2=1.053$, we then arrive at the ensuing compact expression
\begin{equation}\label{eq:PR TM 3}
P_R^{\text{TM}}=\frac{128A_0}{\pi(1+\sqrt{2})^2}\sum_{n=0}^{N-1}\xi_n.
\end{equation}
 On the other hand, the useful mean radiated power is  $P_U^{\text{TM}}=(p_1)^0_1$ and, by virtue of \cref{eq:p_1 k}, we have that
 \begin{equation}\label{eq:P_U}
 P_U^{\text{TM}}=\frac{128}{\pi(1+\sqrt{2})^2} \sum_{n=0}^{N-1}\xi_n^2,
\end{equation}
and by substituting \cref{eq:PR TM 3} and \cref{eq:P_U} into \cref{eq:eta_TMA}, we have 
\begin{equation}\label{eq:efficiency_new3}
\eta_{\text{TMA}}=\frac{\sum_{n=0}^{N-1}\xi_n^2}{A_0\sum_{n=0}^{N-1}\xi_n}.
\end{equation}
Regarding the term $\eta_{\text{BFN}}$, by quantifying $P_R^\text{ST}$ as the total mean transmitted power over the array factor $F^{\text{ST}}(\theta)=\sum_{n=0}^{N-1}\expe{jkz_n\cos\theta}$, we have
\begin{equation}\label{eq:P_R_ST}
 P_{\text{R}}^{\text{ST}}= \int_{0}^{2\pi}\int_{0}^{\pi} |F^{\text{ST}}(\theta)|^2\sin(\theta)\text{d}\theta \text{d}\varphi=4\pi N.
\end{equation} 
By substituting \cref{eq:PR TM 3} and \cref{eq:P_R_ST} into \cref{eq:eta_BFN}, we arrive at 
\begin{equation}\label{eq:eta_BFN final}
\eta_{\text{BFN}}=\frac{32A_0\sum_{n=0}^{N-1}\xi_n}{\pi^2(1+\sqrt{2})^2 N}.
\end{equation}

\section{Numerical Examples}\label{sec:numerical sim}

In this section we examine the behavior of the proposed \ac{TMA} in its two possible configurations: (1) basic (or phased array) mode, with only phase weighting of the array excitations; and (2) full-featured (or beamformer)  mode, with amplitude-phase weighting of the array excitations.
\subsection{Phased Array Mode}\label{subsec:phased array}
Let us consider a uniform linear array with $N=30$ elements spaced $d = \lambda/2$ apart with unitary static excitations treated as the one illustrated in \cref{fig:block diagram n-th element}. We configure the \ac{TMA} by setting $c_n(t)=1$ and, hence, the \ac{SPST} switches are permanently closed. In other words, $k$ is set to zero and $C_{n0}=1,\ n \in \{0, 1,\dots, N-1\}$. Initially, the time-delays $D_n$ are also set to zero, hence the scanning angle of all patterns (see \cref{eq:normalized dynamic excitations}) will be $\theta_{\text{scan}}=90^\circ$. \cref{fig:radiated pattern phased array1}a illustrates that the proposed scheme
is capable of concentrating the radiated power on the desired first harmonic pattern, $|F_1(\theta,t)_1^0|^2$, located at $\omega_c+\omega_0$. The most meaningful unwanted harmonic patterns ---in decreasing order of significance--- are: $|F_2(\theta,t)_7^0|^2$ at $\omega_c-7\omega_0$, $|F_1(\theta,t)_{9}^0|^2$ at $\omega_c + 9\omega_0$, and $|F_2(\theta,t)_{15}^0|^2$ at  $\omega_c - 15\omega_0$.
 It is remarkable that the highest peak level of the unwanted harmonics ($-16.90$\,dB, corresponding to $\omega_c-7\omega_0$) is approximately $4$\,dB below the level of the main secondary lobes ($-13$\,dB) of the desired pattern at $\omega_c+\omega_0$. The subsequent harmonics, at $\omega_c+9\omega_0$ and $\omega_c-15\omega_0$, have peak levels of $-19.08$\,dB and $-23.52$\,dB, respectively. On the other hand, by virtue of \cref{eq:eficiencias}, \cref{eq:efficiency_new3} and \cref{eq:eta_BFN final}, the corresponding efficiencies\footnote{Throughout this paper we express the efficiencies both in natural units (as defined in \cref{eq:efficiency_new3}, \cref{eq:eta_BFN final} and \cref{eq:eficiencias}) and in dB, i.e., as $10\log_{10}(\cdot)$ of the corresponding efficiencies. The latter is more convenient to specify the power losses of the \ac{TMA}.} are $\eta_{\text{TMA}}=0.96$ ($-0.16$\,dB) and $\eta_{\text{BFN}}=0.58$ ($-2.39$\,dB), leading to $\eta=0.56$ ($-2.55$\,dB).

\cref{fig:radiated pattern phased array1}c illustrates the scanning capability of the proposed \ac{TMA} scheme. The $D_n$ values are selected to accomplish a $\theta_{\text{scan}}=70^\circ$ by simply assigning progressive phases to the array elements. Notice that the $D_n$ values (see \cref{{eq:p_1 k}}) have no effect on the efficiency. Also, as we are exclusively performing a phase weighting of the array excitations, the radiated power patterns in Figs.~\ref{fig:radiated pattern phased array1}a and \ref{fig:radiated pattern phased array1}c are, necessarily, uniform. 

\begin{table}[b]
\caption{Normalized pulse durations of the sequences that govern the \ac{SPST} switches shown in \cref{fig:block diagram n-th element}. $\xi_n$ values are provided in \cite{Fondevila2006} where symmetric dynamic excitations are considered in an array of $N=30$ elements. Hence, $n \in \{0, 1,\dots, 29\}$.}
\label{tab: xi_n}
\setlength\tabcolsep{0.2em}
\def\arraystretch{1.1}
\begin{small}
\begin{center}
\begin{tabular}{|c|c|c|c|c|c|c|c|}
	\hline
	{element} & $1$, $28$ & $2$, $27$ & $3$, $26$ & $4$, $25$ & $5$, $24$ & $9$, $20$ & others\\\hline
	{$\xi_n$} & $0.136$ & $0.050$ & $0.953$ & $0.947$ & $0.689$ & $0.926$ & $1$\\
	\hline 
\end{tabular}
\end{center}
\end{small}
\end{table}

\subsection{Beamformer Mode}\label{subsec:beamformer} 
In this mode, the \ac{SPST} switches are governed by the periodic sequences $c_n(t)$. \cref{fig:radiated pattern phased array1}b illustrates the normalized power radiated pattern when $D_n$ are set to zero ($\theta_{\text{scan}}=90^\circ$) and the normalized pulse durations $\xi_n$ of the modulating sequences  $c_n(t)$ are those in \cref{tab: xi_n}. Such time durations were obtained by means of a simulating annealing algorithm as explained in \cite{Fondevila2006}. Under the assumption of symmetric dynamic excitations, for each index $q$, the optimization algorithm is capable (in this example) of setting the \ac{SLL} of the desired pattern ($k=0$) to $-17$\,dB, while the remainder harmonics patterns ($k\neq0$) are kept below $-30$\,dB. Notice that, due to this additional time modulation, apart from the undesired harmonics of the phased array mode  (associated to different indexes $q$ and whose corresponding \ac{SLL} are also set to $-17$\,dB with respect to their maxima), other harmonic patterns are generated, as it was analyzed in \cref{sec:SSB TMA with stair-step pulses model}. Among them, the most significant is $|F_1(\theta,t)_{1}^1|^2$ at  $\omega_c+2\omega_0$ which, as we can observe, is below $-30$\,dB. The price to be paid for the \ac{SLL} improvement is a certain worsening of the \ac{TMA} efficiencies: $\eta_{\text{TMA}}=0.91$ ($-0.41$\,dB) and $\eta_{\text{BFN}}=0.50$ ($-3.01$\,dB), leading to $\eta=0.46$ ($-3.42$\,dB). \cref{fig:radiated pattern phased array1}d illustrates both the scanning and the amplitude reconfiguration capabilities of the proposed \ac{TMA} by showing the normalized power radiated pattern when $D_n$ are selected to accomplish a $\theta_{\text{scan}}=110^\circ$, whereas $c_n(t)$ are the same as those in \cref{fig:radiated pattern phased array1}b.

The results in Figs.~\ref{fig:radiated pattern phased array1}b and \ref{fig:radiated pattern phased array1}d are similar to those presented in \cite{Amin_Yao2015}. Nevertheless, the architecture proposed in this work allows for a maximum signal bandwidth $B_{\text{max}}=8f_0$ (the first unwanted harmonic is $q=-7$), whereas in \cite{Amin_Yao2015}, $B_{\text{max}}=4f_0$, and hence, improving the bandwidth response of the \ac{TMA} by $100$\%.

Notice that the proposed architecture handles two types of time parameters: 
\begin{enumerate}
\item The time delays $D_{n}$ for the modulating signals, which are selected to synthesize progressive phases, i.e., $D_{n}/T_0=n\cos(\theta_{\text{scan}})$, where $\theta_{\text{scan}}$ is the direction of the first positive harmonic beam.

\item The pulse durations, $\xi_n$, corresponding to the on-state duration of the rectangular pulses that govern $c_n(t)$. The values of $\xi_n$ are obtained by means of an optimization algorithm \cite{Fondevila2004a} which selects the Fourier coefficients of $c_n(t)$, $C_{nk}=\xi_n\sinc(k\pi\xi_n)\expe{-jk\pi\xi_n}$, to obtain a radiation diagram for $k=0$ with a given \ac{SLL}, whereas the rest of harmonics ($k\neq0$) are kept below a threshold.
\end{enumerate}

\begin{figure*}[!h]
	\centering
	\includegraphics[width=0.8\linewidth]{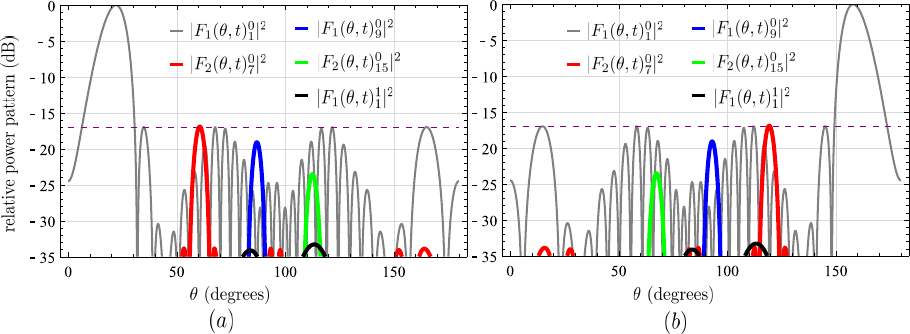}
	\caption{Normalized power radiated pattern showing a \ac{TMA} scanning range of $\pm 68^\circ$ from the broad sight direction. The \ac{TMA} is configured in the beamforming mode with the pulse durations $\xi_n$ in \cref{tab: xi_n}. The threshold level of the unwanted harmonics is not being affected by the beam steering, but the half power beam width of the exploited harmonic widens as it moves away from the center.}
	\label{fig:TMAScanningRange}
\end{figure*}

With respect to the scanning ability of the \ac{TMA}, we observe in \cref{fig:TMAScanningRange} a \ac{TMA} scanning range of $\pm 68$ degrees from the broad sight direction. Although the threshold level of the unwanted harmonics is not being affected by the steering of the beam, the half power beam width of the exploited harmonic widens as it moves away from the center.

\begin{figure}[t]
	\centering
	\includegraphics[width=\linewidth]{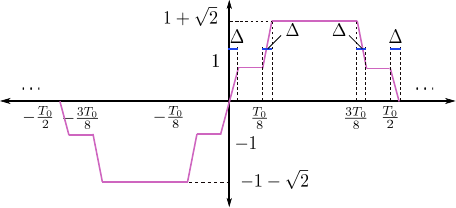}
	\caption{Nonideal stair step pulse. We have considered a rise-fall time of the \ac{SP4T} switches, $\Delta$, and a linear response of the transient, analogously to \cite{Amin_Yao2015,ManeiroCatoira2019a_AWPL}.}
	\label{fig:nonidealStairStepPulse}
\end{figure}

\begin{table}[b]
	\caption{Normalized nonideal pulse durations of the sequences that govern the \ac{SPST} switches shown in \cref{fig:block diagram n-th element}.}
	\label{tab:xi_n_nonidealPulse}
	\begin{small}
		\begin{center}
			\begin{tabular}{|c|c|c|c|c|c|}
				\hline
				{element} & $2$ & $4$ & $5$ & $6$ & $7$ \\\hline
				{$\xi_n$} & $0.063$ & $0.078$ & $0.076$ & $0.063$ & $0.880$ \\
				\hline 
				\hline
				{element} & $14$ & $18$ & $19$ & $20$ & \\\hline
				{$\xi_n$} & $0.962$ & $0.175$ & $0.471$ & $0.977$ &\\
				\hline
			\end{tabular}
		\end{center}
	\end{small}
\end{table}

On the other hand, we have analyzed the impact of the rise-fall times of the switches (see \cref{fig:nonidealStairStepPulse}), and we have realized that there is a time interval (and hence we can select an adequate switch, as in \cite{ManeiroCatoira2019a_AWPL,Amin_Yao2015}), during which the switches can be profitably used to decrease the peak level of the unexploited harmonics, thus improving $\eta_{\text{TMA}}$. However, this is achieved at the expense of reducing the overall time modulation efficiency, $\eta$, and degrading the \ac{TMA} frequency performance because, due to the appearance of new harmonics, the time modulation frequency must be duplicated to faithfully send (receive) signals with the same bandwidth as in the case of considering ideal pulses. 

\begin{figure}[t]
	\centering
	\includegraphics[width=\linewidth]{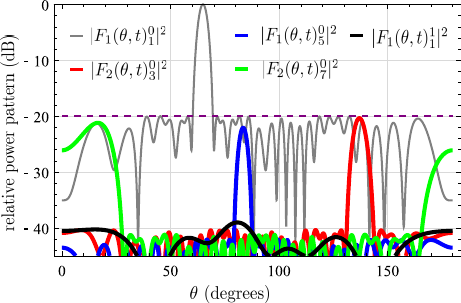}
	\caption{Normalized power radiated pattern of a \ac{TMA} employing nonideal stair step pulses with a rise-fall time equal to $0.06T_0$ and considering the set of $\xi_n$ specified in \cref{tab:xi_n_nonidealPulse} to ensure that $\text{SLL}=-20$\,dB. The threshold for the harmonics is equals to $-20$\,dB, hence improving the efficiency, $\eta_{\text{TMA}}$, but at the expense of degrading the overall efficiency, verifying the trade-off between the \ac{SLL} and the overall efficiency. As a matter of fact, we have obtained: $\eta_{\text{TMA}}=99.0$\% and $\eta_{\text{BFN}}=35.8$\%,  leading to $\eta=35.5$\% ($\eta = -4.45$\,dB).}
	\label{fig:nonidealPulseRadiationPattern}
\end{figure}

We have considered a rise-fall time $0.06 T_0$ together with a set of $\xi_n$ (see \cref{tab:xi_n_nonidealPulse}) to ensure that $\text{SLL}=-20$\,dB and a threshold for the harmonics equal to $-20$\,dB. \cref{fig:nonidealPulseRadiationPattern} shows the corresponding power radiated pattern. With respect to the \ac{TMA} efficiencies we obtain: $\eta_{\text{TMA}}=0.99$ ($\eta_{\text{TMA}}=-0.04$\,dB) and $\eta_{\text{BFN}}=0.36$ ($\eta_{\text{BFN}}=-4.45$\,dB), leading to $\eta=0.35$ ($\eta=-4.49$\,dB). We observe a trade-off between the \ac{SLL} and the overall efficiency of the \ac{TMA} and that, in any case, the total insertion losses introduced by the \ac{TMA} technique are lower than those corresponding to off-the-shelf \acp{VPS} (see \cref{fig:VPS}).

\section{Features of the SSB TMA Beamformer}
In this section we discuss some practial issues regarding the proposed \ac{SSB} \ac{TMA} beamforming technique. In particular, we show the advantages of the proposed scheme with respect to conventional beam scanning antenna systems in relation to the following aspects:


\begin{enumerate}
	\item \textbf{Cost}:
	a common feature to all high frequency reconfigurable devices is the significant increased cost with respect to their non-reconfigurable or fixed counterparts. For example, tunable phase shifters (e.g., \cite{Analog, Custom, Macom, Ommic, Qorvo}) are still an expensive option when compared to fixed broadband phase shifters, which can be manufactured using low-cost printed circuit board technology \cite{Abbosh2011}. A similar reasoning can be applied to fixed RF attenuators \cite{Minicircuits} with respect to variable ones \cite{Qorvo}.
	On the other hand, the use of \ac{SP4T} switches provides a cost-effective solution (see e.g., \cite{Analog}) to carry out the time modulation with stair-step periodic pulses. We could consider other non-switched alternatives, e.g., by using \acp{VGA} \cite{Maneiro2018,Maneiro2017b} or analog multipliers \cite{Maneiro2017b}. 
	
	Nevertheless, the implementation of the time modulation requires such devices be suitable to work in the band of the carrier frequency. As a matter of fact, these devices perform the time modulation by properly processing the transmitted signal at the antenna level, either modulating such a signal in amplitude or multiplying it by a periodic pulse. Therefore, an increase in the carrier frequency translates to a significant cost increase. For instance, if the \ac{TMA} beam scanning operates at a carrier frequency $\omega_c$, it is enough that the \ac{SP4T} switches be suitable to work at the signal bandwidth $B$ regardless of $\omega_c$, whereas the \acp{VGA} or the analog multipliers must work at $\omega_c$. 
	
	\begin{figure}[t]
		\centering
		\includegraphics[width=5cm]{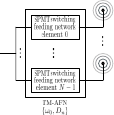}
		\caption{Block diagram of the overall \ac{TM-AFN} in terms of the individual \ac{SPMT} feeding networks of each element (see \cref{fig:block diagram n-th element}), specifying the fundamental frequency of the periodic pulses $\omega_0$, and their time delays $D_n$.}
		\label{fig:TM-AFN}
	\end{figure}
	\begin{figure*}[t]
		\centering
		\includegraphics[width=0.8\linewidth]{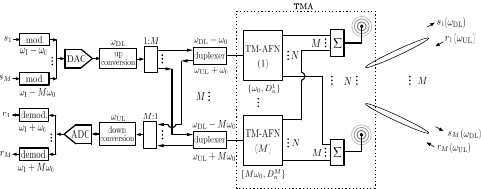}
		\caption{Proposed multibeam transceiver architecture based on \acp{TMA} with \ac{SPMT} switches.}
		\label{fig:multibeamStructure}
	\end{figure*}
	\item \textbf{Complexity}: when we are restricted to a single-beam exploitation, non-switched \ac{TMA} architectures \cite{Maneiro2018, Maneiro2017b} are apparently more complex not only in terms of hardware (especially if the devices have analog control due to the requirements of the \acp{DAC}), but also in terms of software. Whereas \ac{SP4T} switches are governed by the switch-on and switch-off time instants of binary sequences, non-switched \acp{TMA} are capable of constructing, in the digital domain, other complex waveforms, such as the \ac{SWC} pulses \cite{Maneiro2017b} or preprocessed rectangular pulses \cite{Maneiro2018}. Although non-switched \acp{TMA} can perform multibeam harmonic beamforming, they have the serious handicap of their hardware implementation at high frequencies (\acp{VGA} fast enough to follow a wideband signal, e.g., with a bandwidth of $1$\,GHz, or analog multipliers at the millimeter wave band). 
	
	On the other hand, it is well-known that TMA is a bandwidth-limited technique because, in general, the time modulation frequency ($f_0$) must satisfy that $B<f_0$, with $B$ being the signal bandwidth. Such a restriction is necessary to avoid spectral overlapping between the signal replicas located at adjacent harmonics \cite{Maneiro2014}. In our proposal, as the harmonics up to the seventh order are removed, the restriction becomes $B<8f_0$, and hence it is possible to handle signal bandwidths eight times bigger than those handled by multibeam \acp{TMA}. As we have seen in \cref{sec:introduction,sec:mathematicalAnalysis}, the mathematical background of the \acp{TMA} synthesized in this work consists in approximating a pure sinusoid (a single harmonic) by means of stair-step pulses, while keeping the remaining (and inherently generated) harmonics below a threshold. Therefore, the ability of efficiently synthesizing a sum of pure sinusoids with \ac{SPMT} switches for beamforming purposes paves the road for future investigations. Nevertheless, a promising application of the proposed \ac{TMA} feeding network is the transmission (reception) of multiple signals through the same antenna array with different spatial signatures. 
	
	\cref{fig:TM-AFN} shows the block diagram of the overall \ac{TM-AFN} corresponding to the proposed architecture. Notice that such a \ac{TM-AFN} is characterized by the time-modulation frequency $\omega_0$ and the time-delays $D_n$ of the periodic pulses.
	
	Let us consider a transceiver capable of transmitting $M$ linearly modulated digital signals over the downlink frequency $\omega_\mathrm{DL}$ using different spatial signatures as well as capable of receiving $M$ signals with different \acp{DoA} over the uplink frequency $\omega_\mathrm{UL}$. To this end, we consider the multibeam transceiver architecture shown in \cref{fig:multibeamStructure}, which is equipped with $M$ \acp{TM-AFN} such as the one shown in \cref{fig:TM-AFN}.
	
	In the downlink, each complex-valued baseband signal $s_i$, $i \in \{1,\cdots,M\}$, is I/Q modulated at a different intermediate frequency, $\omega_I-i\omega_0$. The I/Q modulated digital signal is then converted to the analog domain by using a single \ac{DAC} with a sampling frequency higher than twice the total bandwidth of the composite signal. The obtained analog signal is up-converted to the downlink frequency, $\omega_\mathrm{DL}$. Next, each individual signal located at $\omega_\mathrm{DL}-i\omega_0$ is filtered out by the corresponding duplexer (equipped with a passband filter centered at such a frequency) before being processed by the corresponding $i$-th \ac{TM-AFN}, which will shift the incoming signal in frequency to the downlink frequency $\omega_\mathrm{DL}$ and will endow it with a spatial signature controlled by $D_n$, while keeping the undesired harmonics below a threshold level. 
	
	With respect to the uplink, the different harmonic patterns are designed to hold spatial orthogonality and, hence, at the output of the $i$-th \ac{TM-AFN}, we have the corresponding received signal $r_i$ located at $\omega_\mathrm{UL}+i\omega_0$. After crossing the corresponding duplexers, all the received signals at $\omega_\mathrm{UL}+i\omega_0$, $i \in \{1,\dots,M\}$, are combined and down-converted to the intermediate frequency, $\omega_I$. The obtained signal is next sampled at a frequency higher than twice the total signal bandwidth in order to be converted to the digital domain by a single ADC. Each digital signal is I/Q decomposed at a different intermediate frequency $\omega_{I}+i\omega_0$ to obtain the baseband complex signal $r_i$. We highlight that, in contrast to antenna arrays with \ac{VPS} architectures \cite[Fig.~10]{Hong2017}, \acp{TMA} employ a single down- (up-) converter and a single ADC (DAC), although requiring a wider bandwidth. 
	
	In sum, when compared to multibeam non-switched \acp{TMA}, the proposed multibeam structure is feasible at high frequencies although requiring a higher complexity. Nevertheless, it is also true that our proposal is less complex than the non-switched alternatives when we are restricted to the exploitation of a single beam. When compared to multibeam arrays based on \acp{VPS} and amplifiers, the proposed \ac{TMA} solution only needs a single down- (up-) converter and a single ADC (DAC).

	\item \textbf{Size}:
	when mobility at high frequencies is indispensable, the size is a crucial aspect. In this sense, \ac{RF} switches and fixed attenuators are available as \ac{MMIC} devices \cite{Minicircuits, Analog}, while fixed phase shifters can be manufactured, for instance, using printed circuit board technology \cite{Abbosh2011}.
	\item \textbf{Performance}: a parameter of paramount importance
	which may determine the applicability of the proposed architecture at high frequencies is the time modulation efficiency $\eta$, theoretically derived in \cref{sec:efficiency} and specifically quantified in \cref{sec:numerical sim}. The counterpart of such an efficiency in standard beamformers based on \acp{VPS} are the insertion losses. As a matter of fact, \cref{fig:VPS} plots a point cloud showing the insertion losses of off-the-shelf \ac{MMIC} digital phase shifters \cite{Aelius,Analog,Custom,Macom,Ommic,Qorvo} for several bands up to $20$\,GHz. The point cloud reveals a certain linearity (Pearson correlation coefficient $r=0.7$) between insertion losses and frequency at these frequencies. The insertion losses (in dB) show the following statistics: range [$2.5$\,dB, $11.0$\,dB], the mean value is $5.91$\,dB, and the standard deviation is $0.81$\,dB. An advantage of the \ac{TMA} scheme proposed in this work is not only that the time modulation efficiency is independent of the carrier frequency, but also that shows competitive values when compared to the \acp{VPS} insertion losses.  	
\end{enumerate}

\begin{figure}[t]
	\centering
	\includegraphics[width=\columnwidth]{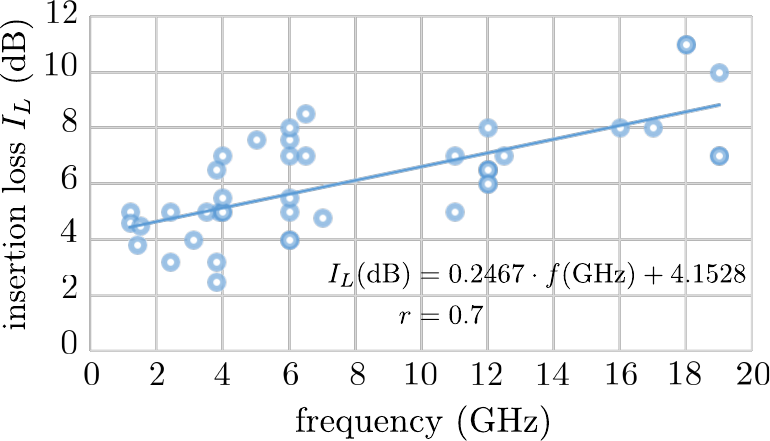}
	\caption{Insertion loss ($I_L$) versus frequency (bands L, S, C, X and Ku) of off-the-shelf \ac{MMIC} digital phase shifters  \cite{Qorvo}, \cite{Analog},\cite{Macom}, \cite{Ommic}, \cite{Custom}, \cite{Aelius}.}
	\label{fig:VPS}
\end{figure}

\section{Conclusion}\label{sec:conclusion}
We have presented a novel \ac{SSB} \ac{TMA} method, based on periodic stair-step pulses, valid for both beam-steering and beamforming purposes. The architecture is equipped with \ac{SP4T} and \ac{SPST} switches together with non-reconfigurable \ac{RF} devices. The proposed \ac{TMA} structure exhibits the following advantages: higher efficiency and flexibility, performance invariant to the carrier frequency, better cost-effectiveness, and small size. Accordingly, such a structure is particularly suitable for the design of multibeam transceivers.

\appendix

\section{Appendixes}\label{Appendix}
\subsection{Derivation of the Expression of the Sign in \cref{eq:Fourier coefficients v(t)}}
\label{Appendix:derivation of sign Vq}
The Triangular numbers 1, 3, 6, 10, 15, $\dots$, given by the formula $T_n=n(n+1)/2$, $n\in \mathbb{N^*}$, have the property that $T_{4k+1}$ and $T_{4k+2}$ are odd and that $T_{4k+3}$ and $T_{4k+4}$ are even for $k\in \mathbb{N}$.  Thus, the expression
\begin{equation}\label{eq:sign triangular}
(-1)^{T_n}=(-1)^{\frac{n(n+1)}{2}}
\end{equation}
alternates its sign according to: $-1$, $-1$, $+1$, $+1$, $-1$, $-1$, $\dots$, and we realize that the sign behavior for $n=1$, $n=2$, $\dots$, corresponds to $q=3$, $q=5$, $\dots$, in the series of \cref{eq:Fourier coefficients v(t)}. Hence, by relating $n$ and $q$ through an arithmetic progression, we have that $q=2n+1$ or, equivalently, $n=(q-1)/2$. By substituting this expression of $n$ in \cref{eq:sign triangular} we finally arrive at the following expression of the sign in \cref{eq:Fourier coefficients v(t)}:
\begin{equation}\label{sign in v(t)}
(-1)^{\frac{(q+1)(q-1)}{8}}.
\end{equation}
\subsection{Derivation of the Sum of the Infinite Series in \cref{eq:PR TM 2}}
\label{subsec:derivation of the sum in eta_P}
The polygamma function is a special function denoted by $\Psi_n(z)$ which is defined as the $(n + 1)$-th derivative of the logarithm of the gamma function $\varGamma(z)$:  
\begin{equation}\label{Psi}
\Psi_n(z)=\frac{d^{n+1}}{dz^{n+1}}\ln\left(\varGamma(z)\right).
\end{equation}
For $n>0$, the polygamma function can be written as \cite[Chapter 6.4, pp. 260-263]{Abramowitz1972} 
\begin{equation}\label{series Psi}
\Psi_n(z)=(-1)^{n+1}n!\sum_{k=0}^{\infty}\frac{1}{(z+k)^{n+1}}.
\end{equation}
In particular, for $n=1$, we have that
\begin{equation}\label{series Psi n=1}
\Psi_1(z)=\sum_{k=0}^{\infty}\frac{1}{(z+k)^2}.
\end{equation}
Hence, the infinite series in the expression of $P_R^{\text{TM}}$  \cref{eq:PR TM 2} will satisfy
\begin{equation}\label{series con polygamma1}
\sum_{k=1}^{\infty}\frac{1}{(8k-7)^2}=\frac{1}{64}\sum_{k=0}^{\infty}\frac{1}{(k+1-\frac{7}{8})^2}=\frac{1}{64}\Psi_1\left(\frac{1}{8}\right),
\end{equation}
and, analogously 
\begin{equation}\label{series con polygamma2}
\sum_{k=1}^{\infty}\frac{1}{(8k-1)^2}=\frac{1}{64}\sum_{k=0}^{\infty}\frac{1}{(k+1-\frac{1}{8})^2}=\frac{1}{64}\Psi_1\left(\frac{7}{8}\right),
\end{equation}
being $\Psi_1(1/8)=65.3881$ and $\Psi_1(7/8)=2.0057$,
and therefore, $\sum_{q\in\Upsilon}1/q^2=1.053$.

\setlength{\bibsep}{5pt}
\bibliographystyle{model5-names}
\bibliography{IEEEabrv,main}

\end{document}